\documentclass[prl,twocolumn,showpacs]{revtex4}
\usepackage{graphicx}
\begin{document}
\title{
   Pairing of Composite Fermions, Laughlin Correlations, \\
   and the Fractional Quantum Hall Hierarchy}
\author{
   Arkadiusz W\'ojs$^{1,2}$, 
   Kyung-Soo Yi$^{1,3}$, and 
   John J. Quinn$^1$}
\affiliation{
   $^1$University of Tennessee, Knoxville, Tennessee 37996, USA\\
   $^2$Wroclaw University of Technology, 50-370 Wroclaw, Poland\\
   $^3$Pusan National University, Pusan 609-735, Korea}
\begin{abstract}
A novel hierarchy of fractional quantum Hall (FQH) states in the 
lowest Landau level (LL) is proposed to explain recently observed 
FQH fractions such as $\nu=5/13$, 3/8, or 4/11. 
Based on the analysis of their interaction pseudopotentials, 
it is argued that the Laughlin quasiparticles (particles/holes 
in a partially filled composite fermion LL) form pairs. 
These pairs are proposed to have Laughlin correlations with
one another and to form condensed states at a sequence of fractions 
which includes all new fractions observed in experiment.
\end{abstract}
\maketitle

\paragraph*{Introduction.}

Pan {\sl et al.} \cite{pan} recently observed fractional quantum 
Hall (FQH) \cite{tsui} minima in the diagonal resistivity of 
a two-dimensional electron gas at novel filling fractions $\nu$ 
of the lowest Landau level (LL).
The new spin-polarized FQH states occur at filling factors outside 
the Jain sequence \cite{jain} of composite fermion (CF) states.
Some of them, as $\nu=4/11$ or 4/13, appear in the Haldane hierarchy 
\cite{haldane} of quasiparticle (QP) condensates, but their 
``hierarchical'' interpretation was questioned \cite{hierarchy} 
because of the specific form of the QP--QP interaction.
Others, such as the $\nu=3/8$ or 3/10 states, do not belong to 
the Haldane hierarchy, and the origin of their incompressibility 
is puzzling in an even more obvious way.
Pan {\sl et al.} take their observations as evidence for residual 
CF--CF interactions, but at the same time they ignore the theoretical 
investigations in which these interactions were studied in detail 
\cite{hierarchy,sitko,lee}.
Consequently, they conclude that the origin of the observed FQH 
states remains unresolved.  
In this letter we propose an explanation for these new states
involving formation of QP pairs which display Laughlin correlations 
with one another.

First, we present briefly the connection between the CF model
(equivalent to the mean-field Chern--Simons transformation) 
\cite{jain} and the QP hierarchy \cite{haldane,hierarchy,sitko} 
to show that the form of the ``residual CF--CF interactions'' is 
neither mysterious nor unknown. 
On the contrary, the pseudopotentials $V(\mathcal{R})$ (defined 
as dependence of pair interaction energy on relative pair angular 
momentum) describing these interactions at short range have been 
calculated \cite{hierarchy,sitko,lee}, and the long-range 
behavior can be readily understood from the nature of the 
(fractionally) charged QP's that interact with one another 
by the (repulsive) Coulomb potential.
It is worth noting the impossibility of deriving $V(\mathcal{R})$ 
from the {\em literally understood} CF picture, where it is 
{\em interpreted} as the difference between Coulomb and gauge 
interactions between fluctuations beyond the mean field.
Second, we recall two simple types of two-body correlations,
Laughlin correlations and pairing, that may occur in an interacting 
system depending on the filling factor $\nu$ and on whether 
$V(\mathcal{R})$ is super- or subharmonic at the relevant range 
\cite{parentage}.
Then, knowing the QP--QP pseudopotential $V_{\rm QP}(\mathcal{R})$, 
we apply the concept of Laughlin condensed states of (bosonic) pairs
(used earlier for the electrons in the $n=1$ LL to describe such FQH 
states as $\nu=5/2$ or 7/3 \cite{fivehalf}) to the particles or holes 
in a partially filled CF LL, i.e., to Laughlin quasielectrons (QE's) 
or quasiholes (QH's).
Finally, we propose the existence of novel hierarchy FQH states 
in which the incompressibility results from the condensation of 
QP pairs (QE$_2$'s or QH$_2$'s) into Laughlin correlated pair states.
The series of FQH states derived from the parent $\nu=1/3$ state 
include all novel fractions reported by Pan {\sl et al.}: $\nu=5/13$, 
3/8, 4/11, and 6/17 for the QE's and $\nu=5/17$, 3/10, 4/13, and 6/19 
for the QH's.

Standard numerical calculations for $N$ electrons are not useful 
for studying the Laughlin correlated pair states because convincing 
results require too large values of $N$.
For a meaningful test, at least three QP pairs need be considered.
For $\nu=4/11$ this occurs for $N=18$ electrons and the total flux 
$2l=45$, which seems beyond reach of exact diagonalization and 
explains the lack of earlier numerical evidence.
In this letter we take advantage of the knowledge of the dominant 
features of $V_{\rm QE}(\mathcal{R})$ to diagonalize two interacting 
QE systems corresponding to $\nu=5/13$ and 3/8.
Although the results confirm nondegenerate ground states and an 
excitation gap, they should be considered as a mere illustration 
in support of our idea, while the most convincing arguments lie 
in the analysis of QP--QP pseudopotentials and in good agreement 
with experiment.
Somewhat similar, detailed calculations were recently carried out 
by Mandal and Jain \cite{mandal} in search of the $\nu=4/11$, 5/13, 
and 6/17 states.
However, these authors did not study systems with the values of 
$(N,2l)$ predicted in our model (and their results did not indicate 
incompressibility).
                              
\paragraph*{QP--QP Pseudopotential.}

The nature of QP correlations depends critically on the 
pseudopotential $V_{\rm QP}(\mathcal{R})$ describing their 
pair interaction energy $V_{\rm QP}$ as a function of relative 
pair angular momentum $\mathcal{R}$.
We have shown earlier \cite{parentage,fivehalf} that the 
correlations are of the Laughlin type (i.e., the particles tend 
to avoid pair states with one or more of the smallest values of 
$\mathcal{R}=1$, 3, \dots) only if $V(\mathcal{R})$ is 
``superharmonic'' at the relevant values of $\mathcal{R}$ for 
a given filling factor $\nu$ (specifically, at $\mathcal{R}=2p-1$ 
for $\nu\sim(2p+1)^{-1}$, where $p=1$, 2, \dots).
Laughlin correlations defined in this way justify reapplication 
of the CF picture to the QP's to select the lowest states of the 
whole many-body spectrum, and lead to the incompressible QP 
``daughter'' states of the standard CF hierarchy \cite{sitko}.
The superharmonic repulsion is defined as one for which $V$ 
decreases more quickly than linearly as a function of the average 
particle--particle separation $\left<r^2\right>$ for the consecutive 
pair eigenstates labeled by $\mathcal{R}$.
In spherical geometry \cite{haldane}, most convenient for finite-size 
calculations, this means that $V$ increases more quickly than linearly 
as a function of $L(L+1)$, i.e., of the squared total pair angular 
momentum $L=2l-\mathcal{R}$, where $l$ is the single-particle angular 
momentum.

\begin{figure}
\resizebox{3.4in}{1.78in}{\includegraphics{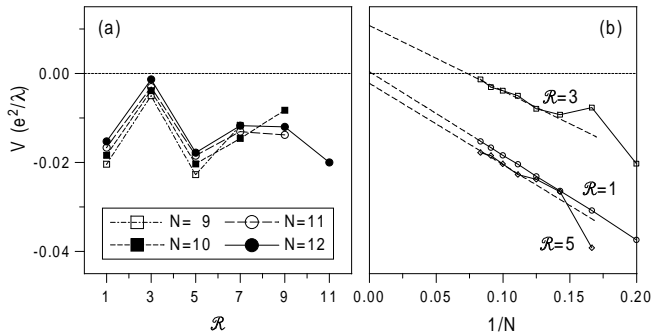}}
\caption{\label{fig1}
   (a) Interaction pseudopotentials $V(\mathcal{R})$ for a pair 
   of QE's of the Laughlin $\nu=1/3$ state calculated for the 
   systems of up to $N=12$ electrons on a sphere.
   (b) Dependence of the leading QE pseudopotential coefficients 
   corresponding to the smallest values of $\mathcal{R}$ on $N^{-1}$.
   Extrapolation to $N^{-1}\rightarrow0$ corresponds to an infinite 
   planar system.} 
\end{figure}

The qualitative behavior of the QP--QP interaction pseudopotential 
$V_{\rm QP}(\mathcal{R})$ at short range is well-known from 
numerical studies of small systems \cite{hierarchy,sitko,lee}.  
In Fig.~\ref{fig1}(a) we compare $V_{\rm QE}(\mathcal{R})$ 
calculated for the systems of $N=9$ to 12 electrons.
The zero of energy is not determined very accurately in finite 
systems, and an extrapolation to large $N$ is needed to restore 
the positive sign of $V_{\rm QE}(\mathcal{R})$, as shown in 
Fig.~\ref{fig1}(b).
However, only the relative values are of importance, since adding 
a constant to $V(\mathcal{R})$ does not affect correlations and only 
shifts the whole many-body spectrum by a (different) constant.
On the other hand, the repulsive character of the QP--QP interaction
and the long-range behavior of $V_{\rm QP}(\mathcal{R})\sim
\mathcal{R}^{-1/2}$ follow from the fact that QP's are charged 
particles (the form of QP charge density affects $V_{\rm QP}$ 
only at short range, comparable to the QP size).

Combining the above arguments, it is clear that the dominant 
features of $V_{\rm QE}$ are the small value at $\mathcal{R}=1$ 
and a strong maximum at $\mathcal{R}=3$.
This result is most apparent in the calculation of Lee {\sl et al.} 
\cite{lee}.
Analogous analysis for the QH's yields maxima at $\mathcal{R}=1$ 
and 5, and nearly vanishing $V_{\rm QH}(3)$.
Importantly, these conclusions do not require such assumptions 
as zero layer thickness $w$ or infinite magnetic field $B$, and 
thus they are readily applicable to the experimental FQH systems.
This is in contrast to the {\em literally understood} CF model
in which the weak ``residual'' CF--CF interactions are said to 
result from partial cancellation of strong Coulomb and gauge 
interactions between the electrons.
These two interactions have very different character and,
for example, depend differently on $w$ or $B$ \cite{parentage}.
This prevents drawing conclusions about even the sign of $V_{\rm QP}$ 
from the CF model alone (which nevertheless remains a useful picture 
for certain other properties of Laughlin QP's).
Let us also note that while features in $V_{\rm QP}(\mathcal{R})$
reflect the internal structure of QP's interacting through bare 
Coulomb potential $V(r)\sim r^{-1}$, it may be convenient to picture 
QP's as point particles interacting through an appropriate model 
potential, for example $V_{\rm QE}(r)\sim (a^2+(r-b)^2)^{-1/2}$ 
with a maximum at $r=b$ corresponding to $\mathcal{R}=3$.

\paragraph*{QP Pairing.}

It is evident that because $V_{\rm QE}(3)>V_{\rm QE}(1)$, the QE 
system does not support Laughlin correlations.  
Instead, we expect that at least some of the QE's will form pairs 
(QE$_2$) at $\mathcal{R}=1$.
A paired state would be characterized by a greatly reduced fractional 
parentage $\mathcal{G}$ \cite{parentage} from the strongly repulsive 
$\mathcal{R}=3$ state compared to the Laughlin correlated state, and 
have lower total interaction energy $E={1\over2}N(N-1)\sum_\mathcal{R}
\mathcal{G}(\mathcal{R})V(\mathcal{R})$.
Let us stress that such pairing is not a result of some attractive
QE--QE interaction, but due to an obvious tendency to avoid the most 
strongly repulsive $\mathcal{R}=3$ pair state.
At sufficiently high QE density this can only be achieved by having 
significant $\mathcal{G}(1)$, which can be interpreted as pairing 
into the QE$_2$ molecules.
By analogy, the QH pairing is expected in the low-energy 
$\mathcal{R}=3$ state.
The range of QP filling factors $\nu_{\rm QP}$ at which pairing 
can be considered is limited by the condition that the separation 
between the closest pairs must exceed the pair size.
While for the QE pairs with $\mathcal{R}=1$ this is satisfied at 
any $\nu_{\rm QE}<1$, the QH pairing with $\mathcal{R}=3$ can 
only occur at $\nu_{\rm QH}<1/3$.

Because of the limited knowledge of $V_{\rm QP}$ at intermediate 
$\mathcal{R}$, we cannot completely preclude pairing into larger 
molecules (e.g. QE$_2$'s with $\mathcal{R}=5$ or QH$_2$'s with 
$\mathcal{R}=7$) that might occur at approprietly lower values 
of $\nu_{\rm QP}$.
We also realize that whether all or only some of the QP's form 
pairs might depend on $\nu_{\rm QP}$, but here we concentrate 
on the simpler, complete-pairing scenario, discussed in the context 
of electrons partially filling the $n=1$ LL \cite{fivehalf}.

\paragraph*{Laughlin Correlations Between Pairs.}

Having established that the QP fluid consists of QP$_2$ molecules, 
the QP$_2$--QP$_2$ interactions need be studied to understand 
correlations.
Here, the QP$_2$'s will be treated as bosons, although in 
two dimensions they can be easily converted to fermions by 
a transformation consisting of attachment of one flux quantum.
The QP$_2$--QP$_2$ interaction is described by an effective 
pseudopotential $V_{{\rm QP}_2}(\mathcal{R})$ that includes 
the correlation effects caused by the fact that the two-pair 
wavefunction must be symmetric under exchange of the whole 
QP$_2$ bosons and at the same time antisymmetric under exchange 
of any pair of the QP fermions.
This problem is analogous to that of interaction between the 
electron pairs in the $n=1$ LL \cite{fivehalf}.

Although we do not know $V_{{\rm QP}_2}(\mathcal{R})$ accurately,
we expect that since it is due to the repulsion between the QP's 
that belong to different QP$_2$ pairs, it might be superharmonic 
at the range corresponding to the QP$_2$--QP$_2$ separation.
For example, the tendency to avoid the pair-QE state at 
$\mathcal{R}=3$ that is responsible for QE pairing is also expected 
to cause spatial separation of the QE$_2$'s in order to reduce the 
contribution to the parentage $\mathcal{G}(3)$ coming from pairs of 
QE's that belong to different QE$_2$'s.
Our numerical results for four QE's (calculations of the expectation 
value of the interaction energy for a model QE--QE pseudopotential 
in the eigenstates of a pairing interaction Hamiltonian) seem to 
support this idea.
However, in contrast to the $n=1$ electron LL \cite{fivehalf},
the lack of accurate data for $V_{\rm QP}$ at the intermediate 
range makes such calculations uncertain.

\paragraph*{Condensed Pair States.}

The assumption of Laughlin correlations between the QP$_2$ bosons 
for the relevant values of $\nu_{\rm QP}$ implies the sequence of 
Laughlin condensed QP$_2$ states that can be conveniently described 
using the ``composite boson'' (CB) model \cite{fivehalf}.
Let us use spherical geometry and consider the system of $N_1$ 
fermions (QP's) each with (integral or half-integral) angular 
momentum $l_1$ (i.e., in a LL of degeneracy $g_1=2l_1+1$).
Neglecting the finite-size corrections, this corresponds to the 
filling factor $\nu_1=N_1/g_1$.
Let the fermions form $N_2={1\over2}N_1$ bosonic pairs each with 
angular momentum $l_2=2l_1-\mathcal{R}_1$, where $\mathcal{R}_1$ 
is an odd integer.
The filling factor for the system of pairs, defined as $\nu_2=N_2/g_2$ 
where $g_2=2l_2+1$, equals to $\nu_2={1\over4}\nu_1$.
The allowed states of two bosonic pairs are labeled by total 
angular momentum $L=2l_2-\mathcal{R}_2$, where $\mathcal{R}_2$ is an
even integer.
Of all even values of $\mathcal{R}_2$, the lowest few are not allowed
because of the Pauli exclusion principle applied to the individual 
fermions.
The condition that the two-fermion states with relative angular 
momentum smaller than $\mathcal{R}_1$ are forbidden is equivalent 
to the elimination of the states with $\mathcal{R}_2\le4\mathcal{R}_1$ 
from the two-boson Hilbert space.
Such a ``hard core'' can be accounted for by a CB transformation
with $4\mathcal{R}_1$ magnetic flux quanta attached to each boson
\cite{theorem}.
This gives the effective CB angular momentum $l_2^*=l_2-2\mathcal{R}_1
(N_2-1)$, effective LL degeneracy $g_2^*=g_2-4\mathcal{R}_1(N_2-1)$, 
and effective filling factor $\nu_2^*=(\nu_2^{-1}-4\mathcal{R}_1)^{-1}$.

The CB's defined in this way condense into their only allowed
$l_2^*=0$ state ($\nu_2^*=\infty$) when the corresponding fermion 
system has the maximum density at which pairing is still possible,
$\nu_1=\mathcal{R}_1^{-1}$.
At lower filling factors, the CB LL is degenerate and the spectrum
of all allowed states of the $N_2$ CB's represents the spectrum of 
the corresponding paired fermion system.
In particular, using the assumption of the superharmonic form of 
boson--boson repulsion, condensed CB states are expected at 
a series of Laughlin filling factors $\nu_2^*=(2q)^{-1}$.
Here, $2q$ is an even integer corresponding to the number of
additional magnetic flux quanta attached to each CB in a subsequent
CB transformation, $l_2^*\rightarrow l_2^{**}=l_2^*-q(N_2-1)$,
to describe Laughlin correlations between the original CB's of
angular momentum $l_2^*$.
From the relation between the fermion and CB filling factors,
$\nu_1^{-1}=(4\nu_2^*)^{-1}+\mathcal{R}_1$, we find the following 
sequence of fractions corresponding to the Laughlin condensed pair 
states, $\nu_1^{-1}=q/2+\mathcal{R}_1$.
Finally, we set  $\mathcal{R}_1=1$ for the QE's and $\mathcal{R}_1=3$ 
for the QH's, and use the hierarchy equation \cite{hierarchy}, 
$\nu^{-1}=2p+(1\pm\nu_{\rm QP})^{-1}$, to calculate the following 
sequences of electron filling factors, $\nu$, derived from the parent 
$\nu=(2p+1)^{-1}$ state
\begin{equation}             
   \nu^{-1}=2p+1\pm(2+q/2)^{-1},
\end{equation}
where ``$-$'' corresponds to the QE's and ``$+$'' to the QH's.
Remarkably, all the fractions reported by Pan {\sl et al.} are
among those predicted for the $\nu=1/3$ parent and listed in 
Tab.~\ref{tab1}. 
\begin{table}
\caption{\label{tab1}
   The $\nu=1/3$ hierarchy of Laughlin states of QP pairs.
   Fractions in boldface have been reported in Ref.~\cite{pan}.}
\begin{ruledtabular}
\begin{tabular}{lcccccccc}
   $q$&1&2&3&4&5&6&7&8\\
\hline
   $\nu_{\rm QE}$&2/3&1/2&2/5&1/3&2/7&1/4&2/9&1/5\\ 
   $\nu$&{\bf5/13}&{\bf3/8}&7/19&{\bf4/11}&9/25&5/14&11/31&{\bf6/17}\\
\hline
   $\nu_{\rm QH}$&2/7&1/4&2/9&1/5&2/11&1/6&2/13&1/7\\
   $\nu$&{\bf5/17}&{\bf3/10}&7/23&{\bf4/13}&9/29&5/16&11/35&{\bf6/19}\\
\end{tabular}
\end{ruledtabular}
\end{table}
Note also that the same values of $q=1$, 2, 4, and 8 describe
both observed QE and QH states.
This indicates similarity of the QE--QE and QH--QH pseudopotentials 
and suggests that both $V_{{\rm QE}_2}$ and $V_{{\rm QH}_2}$ may be
superharmonic only at the corresponding four values of $\mathcal{R}$
(in such case, remaining fractions of Tab.~\ref{tab1} could not be 
observed even in most ideal samples).
Another possibility is only partial pairing of QP's at some of
the filling factors.
Since the pseudopotentials for QP's of different Laughlin states 
are quite similar \cite{hierarchy}, let us also list the fractions 
of the $\nu=1/5$ hierarchy corresponding to $q=1$, 2, 4, and 8:
$\nu=5/23$, 3/14, 4/19, and 6/29 for the QE's and
$\nu=5/27$, 3/16, 4/21, and 6/31 for the QH's.

\paragraph*{Numerical Results.}

As an illustration, we performed exact-diagonalization calculations 
on a sphere for systems of QE's interacting through a model 
pseudopotential with only one coefficient, $V_{\rm QE}(3)=1$.
Because a harmonic pseudopotential does not change the many-body 
eigenstates and only shifts the energy spectrum by a term that is 
linear in $L(L+1)$, such choice of $V_{\rm QE}$ means that we neglect 
all anharmonic contributions to the QE--QE interaction except for 
the strongest one at $\mathcal{R}=3$ \cite{hierarchy,sitko,lee}.

\begin{figure}
\resizebox{3.4in}{1.78in}{\includegraphics{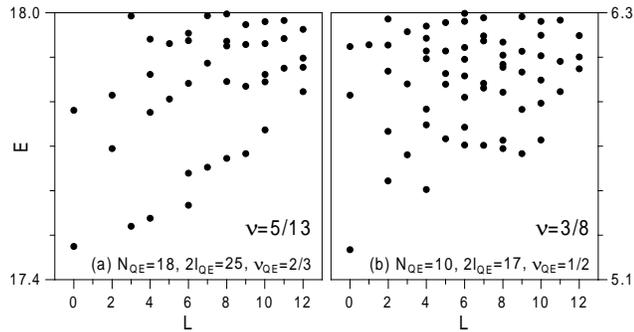}}
\caption{\label{fig2}
   Low energy spectra (energy $E$ as a function of total angular 
   momentum $L$) of 16 QE's at $2l_{\rm QE}=25$ corresponding to 
   $\nu_{\rm QE}=2/3$ and $\nu=5/13$ (a) and 10 QE's at 
   $2l_{\rm QE}=17$ corresponding to $\nu_{\rm QE}=1/2$ and 
   $\nu=3/8$ (b).}
\end{figure}

In a numerical calculation it is important to account for the 
finite-size corrections when calculating $N_1$ and $l_1$ 
corresponding to a given $\nu$.
As an example for $\nu=5/13$ corresponding to $\nu_{\rm QE}\equiv
\nu_1=2/3$ and $q=1$, we pick $N_{\rm QE}\equiv N_1=18$ and calculate 
$N_2={1\over2}N_1=9$, $l_2^*=q(N_2-1)=8$, $l_2=l_2^*+2\mathcal{R}_1
(N_2-1)=24$, and finally $2l_{\rm QE}\equiv 2l_1=l_2+1=25$.
This QE system represents $N=N_1+2l_1-1=42$ electrons at 
$2l=2(l_1-1)+2p(N-1)=105$.
For $\nu=3/8$ corresponding to $\nu_1=1/2$ and $q=2$, we pick $N_1=10$ 
and calculate $N_2=5$, $l_2^*=8$, $l_2=16$, and finally $2l_1=17$.
This system represents $N=26$ and $2l=65$.
Let us note that the values of $(N,2l)$ used here differ from 
those of Ref.~\cite{mandal} where pairing between QP's was not 
considered.
The energy spectra for these two systems are shown in Fig.~\ref{fig2}.
They both show a nondegenerate ($L=0$) ground state and a finite gap.

The very simple QE--QE pseudopotential we used did not give the 
expected $L=0$ ground state in every case we tested.  
However, the lowest $L=0$ state was always the lowest state for small 
values of $L$ which suggests that it might become an absolute ground 
state if the appropriate harmonic term was included in $V_{\rm QE}$.
Further numerical studies of the energy spectra and of the 
coefficients of fractional parentage $\mathcal{G}$ may clarify 
whether this is caused by the use of an oversimplified model 
pseudopotential or perhaps only partial QE pairing at 
$\nu_{\rm QE}<1/2$.

\paragraph*{Conclusions.}

We have studied the QP--QP interactions leading to novel 
spin-polarized FQH states in the lowest LL.
Using the knowledge of QP--QP pseudopotentials and a general 
dependence of the form of correlations on the super- or 
subharmonic behavior of the pseudopotential, we have shown 
that QP's form pairs over a certain range of filling factor 
$\nu_{\rm QP}$.
Then, we argued that the correlations between the QP pairs 
should be of Laughlin type and proposed a hierarchy of 
condensed paired QP states, in analogy to the paired states 
in the excited electron LL studied earlier.
The proposed hierarchy of fractions agrees remarkably well 
with the recent experiment of Pan {\sl et al.} \cite{pan}.
However, more detailed calculations are needed to understand 
why only some of the predicted states were observed.
In particular, further studies might verify the possibility 
of only partial pairing and formation of mixed liquids 
containing both paired and unpaired QP's at filling factors
corresponding to $q>2$.
Also, the whole $\nu=1/5$ hierarchy is yet to be confirmed 
experimentally.
Finally, stability of the proposed states against spin excitations 
at low Zeeman energy needs to be tested, as a partially polarized 
$\nu=4/11$ state was suggested by Park and Jain \cite{park}, 
and its particle-hole conjugate, $\nu=7/11$, appears unpolarized 
in experiment \cite{pan}.

\acknowledgments

This work was supported by Grant DE-FG 02-97ER45657 of the Materials
Science Program -- Basic Energy Sciences of the U.S. Dept.\ of Energy.  
AW acknowledges support from Grant 2P03B02424 of the Polish KBN.
KSY acknowledges support from Grant R14-2002-029-01002-0 of the KOSEF.
The authors thank Jennifer J. Quinn and Josef Tobiska for helpful 
discussions.

\end{document}